\RequirePackage{fix-cm}
\documentclass[twocolumn]{svjour3}          % twocolumn
\smartqed  % flush right qed marks, e.g. at end of proof
\usepackage{graphicx}
\usepackage{amsmath}
\usepackage{amsfonts}
\usepackage{multirow}
\usepackage{fullpage}
\usepackage{algorithmic}
\usepackage{natbib}
\usepackage{setspace}
\usepackage{bm}
\usepackage{graphicx}
\usepackage{hyperref}
\usepackage{booktabs}
\usepackage{bbm}
\usepackage{enumerate}
\usepackage{color}
\usepackage{caption}
\usepackage{subcaption}
\allowdisplaybreaks

\bibpunct{(}{)}{;}{a}{,}{,}

\DeclareMathOperator*{\argmax}{arg\,max}
\newcommand{\dif}{\mathrm{d}}

\DeclareGraphicsExtensions{.eps,.ps,.pdf}

\begin{document}

\title{Copula Modeling for Data with Ties%\thanks{Grants or other notes
}

\titlerunning{Copula for Data with Ties}        % if too long for running head

\author{Yan Li \and Yang Li \and Yichen Qin \and
Jun Yan}

\authorrunning{Y. Li, Y. Li, Y. Qin, and J. Yan} % if too long for running head

\institute{Yan Li \at
              University of Connecticut\\
              \email{yan.4.li@uconn.edu}           %  \\
           \and
           Yang Li (Corresponding author) \at
              Renmin University of China\\
              \email{yang.li@ruc.edu.cn}
           \and
           Yichen Qin \at
              University of Cincinnati \\
              \email{qiuyn@ucmail.uc.edu}
           \and
           Jun Yan \at
              University of Connecticut \\
              \email{jun.yan@uconn.edu}
}

\date{Received: December 2016 / Accepted: date}

\maketitle

\begin{abstract}
Copula modeling has gained much attention in many fields recently with the
advantage of separating dependence structure from marginal distributions.
In real data, however, serious ties are often present in one or multiple
margins, which cause problems to many rank-based statistical methods
developed under the assumption of continuous data with no ties.
Simple methods such as breaking the ties at random or using average rank
introduce independence into the data and, hence, lead to biased estimation.
We propose an estimation method that treats the ranks of tied data
as being interval censored and maximizes a pseudo-likelihood
based on interval censored pseudo-observations.
A parametric bootstrap procedure that preserves the observed tied
ranks in the data is adapted to assess the estimation uncertainty
and perform goodness-of-fit tests.
The proposed approach is shown to be very competitive in comparison
to the simple treatments in a large scale simulation study.
Application to a bivariate insurance data illustrates the methodology.

%% need to be in alphabetical order

\end{abstract}

\section{Introduction}
\label{sec:intro}

Multivariate modeling based on copulas has been extensively applied
in many fields such as finance \citep[e.g.,][]{Mackenzie2014}, actuarial
science \citep[e.g.,][]{You2014}, hydrology \citep[e.g.,][]{Parent2014},
public heath \citep[e.g.,][]{Hu2014}, and so on.
An important advantage of such models is that the dependence structure
of a multivariate distribution is separated from its marginal distributions.
The most popular approach to copula modeling is rank-based, which
does not specify the parametric form of the marginal distributions
\citep[e.g.,][]{Genest1995, GenestGhoudiRémillard2007}.
Under the assumption of continuous marginal distributions, no ties
are expected from the observed data so the ranks are unique.
In many applications, however, ties are often present in one or multiple
margins due to precision limit and rounding in observed data.
For example, two variables, loss and expenses, in an insurance
application \citep{FreesValdez1998} from 1466 uncensored claims
have only 541 and 1401 unique values, respectively.
Presence of ties may have significant effect on the accuracy of parameter
estimation and statistical testing for copulas due to the rank-based
method \citep{Kojadinovic2010, Genest2011, Kojadinovic2016}.
Ties may occur in practice due to two major reasons:
precision/rounding issue and the discontinuity of true marginal models.
We assume that the true marginal distributions are continuous, so we only
consider the first situation, where ties cause information loss.

Handling data with ties in copula modeling has not been fully studied.
Discarding the ties is obviously not desirable because it throws data away.
In rank-based methods, naive approaches are to use average rank or to break
the ties at random multiple times and summarize the multi-data results.
\citet{Kojadinovic2010} compared the two naive methods using the bivariate
insurance data from \citet{FreesValdez1998}: both methods give similar
parameter estimates, but in goodness-of-fit test, using average rank rejects
the Gumbel copula which fits well the data as indicated by results from 100
replicates from breaking ties at random.
Conceptually, both naive methods introduce independence into the data.
Neither of them accounts for the dependence information hidden in the tied
data, their estimation may be biased, especially when the dependence is
strong, and goodness-of-fit tests will not hold their sizes by overly
rejecting the null hypothesis when the null hypothesis is true.
\citet{Pappada2016} proposed two randomization strategies beyond the naive
independence randomization: co-monotone and mixed randomization
(which mixes the co-monotonicity and the independence via some weight).
Nonetheless, the co-monotone randomization introduces perfect dependence into
the data, and the mixed randomization alters the distribution of the data,
albeit less severely.

We propose to handle tied data by treating their ranks as being
interval censored and using ideas for interval censored data
from survival analysis \citep[e.g.,][]{Sun2007, Chen2012}.
For bivariate data, each pair of observation falls into four categories:
both observed, exactly one or the other observed, or both censored.
Interval censored pseudo-observations can be used to construct a
pseudo-likelihood, which can be maximized to obtain point estimates.
To make inferences, the standard parametric bootstrap would not capture the
variation in the estimation because bootstrap samples contain no ties.
We propose a parametric bootstrap procedure that preserves the ties in the
observed data in each bootstrap sample inspired by \citet{Bucher2015}.
The same bootstrap procedure can be used in goodness-of-fit tests to assess
the significance of a wide class of testing statistics constructed from the
goodness-of-fit empirical process \citep{Genest2009, Kojadinovic2011}.
In a large scale simulation study, the point and interval estimation were
shown to be unbiased and provide valid uncertainty measures, respectively;
the goodness-of-fit tests maintained their sizes and have substantial power.

The rest of this article is organized as follows.
The proposed method is described with detail in Section~\ref{sec:meth}.
A large scale numerical study is reported in Section~\ref{sec:simu}.
The insurance data is used to illustrate the method in Section~\ref{sec:reals}.
A discussion concludes in Section~\ref{sec:disc}.

\section{Methodology}
\label{sec:meth}

\subsection{Interval Censored Pseudo-Observations}\label{sec:framework}

Let $(X, Y)$ be a continuous random vector with marginal distribution
functions $F$ and $G$, and joint distribution function $H$.
By Sklar's theorem \citep{Sklar1959}, there is a unique copula
$C: [0, 1]^{2} \rightarrow [0, 1]$ such that
\[
H(x, y) = C\big(F(x),  G(y)\big).
\]
The copula $C$ completely characterizes the dependence structure in $H$.
This representation suggests that the dependence structure can be
separated from the marginal distributions in multivariate modeling.
Let $(X_i, Y_i)$, $i = 1, \ldots, n$ be a random sample from $H$.
Often, the marginal distributions are modeled by their empirical
distributions and the copula is modeled parametrically, leading to a
semiparametric inference in multivariate modeling \citep{Genest1995}.
This approach avoids the bias in copula estimation caused by
misspecified marginal distributions \citep{Kim2007}.

Continuous data have no ties and no ambiguity in ranks.
Let $\hat F_n$ and $\hat G_n$ be the empirical
distribution functions of $F$ and $G$, respectively.
Pseudo-observations $U_i$ and $V_i$ are simply
$\hat F_n (X_i)$ and $\hat G_n (Y_i)$
rescaled by a constant $n / (n + 1)$ to avoid evaluation of the
copula density on the edges of unit square ending at $(1, 1)$.
That is,
\begin{align}
\label{eq:pobs}
(U_i, V_i) = \bigg( \frac{n}{n+1} \hat F_n (X_i),
\frac{n}{n+1} \hat G_n (Y_i) \bigg),
\end{align}
for $i = 1, \ldots, n$.
Without ties, the pseudo-observations at each margin have
jumps of size $1 / (n + 1)$.

The pseudo-likelihood estimator of $\theta$ is constructed from the
margin-free pseudo-observations \citep{Genest1995}:
\[
\hat\theta_n = \argmax_{\theta \in \Theta} \sum_{i=1}^n
\log c\big(U_i, V_i; \theta),
\]
where $c(\cdot, \cdot; \theta)$ is the density of $C$ with parameter
vector $\theta$ and parameter space $\Theta$.

In practice, ties are commonly observed due to rounding
or lack of precision in measurements, which makes ranks and
pseudo-observations not fully observed but interval censored.
An interval censored observation is a data point that is known to be
somewhere between two values but the exact value is unknown.
For illustration, consider a toy example of 9 observations where the
sorted pseudo observations of $X$ from~\eqref{eq:pobs} are
\begin{equation}
\label{eq:toy}
(U_1, \ldots, U_9) = (1, 2, 5, 5, 5, 6, 8, 8, 9) / 10.
\end{equation}
In this example, there are ties in the 3rd, 4th, and 5th pseudo-observations
and in the 7th and 8th pseudo-observations.
If average ranks (also known as mid-ranks) are used, they will be 4 and 7.5.
Handling ties by their average ranks invalidates the parametric bootstrap
method because no ties would be in bootstrap samples, and the distribution of
the many test statistics is not well approximated \citep{Kojadinovic2016}.
Breaking the ties at random gives many possibilities of untied data,
whose results could be summarized \citep{Kojadinovic2010}.
As shown in our simulation study, however, breaking the ties at random
can lead to bias in copula estimation when the dependence is high, which is
expected because it introduces independence into the data, ignoring the
dependence among the interval censored pseudo-observations.

We propose to use the concept of interval censored data from
survival analysis to handle tied data in copula estimation.
In particular, we define upper and lower boundaries of
pseudo-observations, respectively, as
\begin{align*}
(\overline{U}_i, \overline{V}_i)
&=\left( \frac{n \hat F_n (X_i)}{ n+1 },
\frac{ n \hat G_n (Y_i) }{ n+1 } \right),\\
(\underline{U}_i, \underline{V}_i)
&=\left( \frac{ n\hat F_n (X_i-)+1 }{ n+1 } ,
\frac{ n \hat G_n (Y_i-)+1}{ n+1 } \right).
\end{align*}
where $\hat F_n (x-)$ and $\hat G_n (y-)$ are the left limit
of $\hat F_n$ and $\hat G_n$ at $x$ and $y$, respectively.
Note that the upper bounds are the same as $(U_i, V_i)$.
If $X_i$ (or $Y_i$) is a tied observation, then its pseudo observation
$U_i$ (or $V_i$) is interval censored by $[\underline{U}_i, \overline{U}_i]$
(or $[\underline{V}_i, \overline{V}_i]$).  If $X_i$ (or $Y_i$) is not
a tied observation, the interval reduces to a single value, i.e.,
$\underline{U}_i= \overline{U}_i=U_i$ (or
$\underline{V}_i= \overline{V}_i=V_i$).

\subsection{Pseudo-Likelihood Estimator}\label{sec:est}

The observation $(U_i, V_i)$'s contribution to the pseudo likelihood,
$L_i(\theta)$, depends on the censoring pattern on the two margins.
There are four cases.
\begin{enumerate}[(1)]
\item
If $\underline{U}_i < \overline{U}_i $ and $\underline{V}_i < \overline{V}_i$
(i.e., the observation is tied observation in both margins), then
$L_i(\theta)$ is
\begin{align*}
% = \int_{ \underline{U}_i }^{ \overline{U}_i } \int_{ \underline{V}_i }^{
%   \overline{V}_i } c_{\theta} ( u, v; \theta ) \dif u \dif v \newline
& C_{\theta} ( \overline{U}_i, \overline{V}_i ) -
C_{\theta} ( \overline{U}_i, \underline{V}_i ) - \\
& \quad
C_{\theta}( \underline{U}_i, \overline{V}_i ) +
C_{\theta}( \underline{U}_i, \underline{V}_i ).
\end{align*}

\item
If $\underline{U}_i < \overline{U}_i $ and
$\overline{V}_i = \underline{V}_i = V_i$
(i.e., the observation is a tied observation only in $X$), then
$L_i(\theta)$ is
\[
\frac{ \partial C_{\theta} ( u, v ) }{ \partial v } \bigg|_{u =
\overline{U}_i, v = V_i }- \frac{ \partial C_{\theta} ( u, v ) }{
\partial v } \bigg|_{u = \underline{U}_i, v = V_i}.
\]

\item
If $U_i = \underline{U}_i = \overline{U}_i $ and
$\overline{V}_i < \underline{V}_i$
(i.e., the observation is a tied observation only in $Y$), then
$L_i(\theta)$ is
\[
\frac{ \partial C_{\theta} ( u, v ) }{ \partial u }
\bigg|_{ u = U_i, v = \overline{V}_i}
-
\frac{ \partial C_{\theta}( u, v ) }{ \partial u }
\bigg|_{ u = U_i, v = \underline{V}_i }.
\]

\item
If $\underline{U}_i = \overline{U}_i  = U_i$ and
$\overline{V}_i=\underline{V}_i = V_i$
(i.e., the observation is not tied in either margin), then
$L_i(\theta) = c(U_i, V_i;\theta)$.
\end{enumerate}

The adjusted pseudo-likelihood function under interval censoring is
\[
\mathcal{L}(\theta) = \sum_{i=1}^n \log L_i(\theta).
\]
The maximum pseudo-likelihood estimation (MPLE) of $\theta$ is then
\begin{equation}
\label{eq:mple}
\hat{\theta}_n =\argmax_{\theta \in \Theta}\mathcal{L}(\theta).
\end{equation}
This estimator reduces to the traditional MPLE
when neither margin has tied observations.
For implementation, we need partial derivatives of the copula
in addition to the distribution and density functions.
Expressions of these partial derivatives for commonly used copulas
are available from R package \texttt{copula} \citep{Rpkg:copula}.

\subsection{Confidence Interval Estimation}
\label{sec:conf}

The asymptotic properties of the pseudo-likelihood estimator are challenging
to establish due to the inclusion of interval censored pseudo-observations.
We resort to bootstrap for confidence intervals, but a plain vanilla
parametric bootstrap procedure would not work in this case because no ties
would be present if bootstrap samples are generated from the fitted copulas.
The parametric bootstrap procedure needs to be modified so that the ties in
the observed data are somehow preserved in each of the bootstrap samples
in order to sufficiently capture the uncertainty in parameter estimation.

Given a sample generated from the fitted copula, which contains no ties,
we introduce ties into the sample such that at each margin the ties
in the observed data are reproduced in the bootstrap sample.
Let $\tilde F_n$ and $\tilde G_n$ be the empirical distribution of the
observed pseudo-observations $U_i$'s and $V_i$'s, respectively, i.e.,
$\tilde F_n(u)=\sum_{i=1}^{n} \mathbbm{1}(U_i \leq u)/n$ and
$\tilde G_n(v)=\sum_{i=1}^{n} \mathbbm{1}(V_i \leq v)/n$.
When ties are present, $\tilde F_n$ and $\tilde G_n$ have jumps
of sizes greater than $1/n$.
Let $U_i^{(b)}$'s and $V_i^{(b)}$'s be the pseudo-observations from a
bootstrap sample, which have no ties, generated from the fitted copula.
Ties are introduced into to $U_i^{(b)}$'s and $V_i^{(b)}$'s by applying the
corresponding quantile functions $\tilde F_n^{-1}$ and $\tilde G_n^{-1}$
of $\tilde F_n$ and $\tilde G_n$ to $U_i^{(b)}$'s and $V_i^{(b)}$'s,
respectively:
\begin{equation}
\begin{split}
\label{eq:preserve_ties}
&\left(U_i^{(b)},V_i^{(b)}\right)=\left(\tilde F_n^{-1}(U_i^{(b)}), \tilde
G_n^{-1} (V_i^{(b)})\right),\\
&i = 1, \ldots, n,
\end{split}
\end{equation}
where $\tilde F_n^{-1}(y)=\inf\{u: \tilde F_n(u) \geq y \}$.
After this transformation, $U_i^{(b)}$'s and $V_i^{(b)}$'s are tie-adjusted
bootstrap pseudo-observations whose marginal empirical distributions are the
same as those of $U_i$'s and $V_i$'s, respectively \citep{Bucher2015}.
Note that the joint empirical distribution of $(U_i^{(b)}, V_i^{(b)})$,
however, is not the same as that of $(U_i, V_i)$, which is the source
of variation of the bootstrap sample.

After ties are introduced, we can further obtain the upper and lower
boundaries of the pseudo-observations of $U_i^{(b)}$'s and $V_i^{(b)}$'s,
\begin{align*}
\left(\overline{U}_i^{(b)},\overline{V}_i^{(b)} \right)
&= \left(U_i^{(b)}, V_i^{(b)} \right),\\
\left(\underline{U}_i^{(b)},\underline{V}_i^{(b)} \right)
&= \Big(\tilde F_n^{-1}(U_i^{(b)}-)+\frac{1}{n+1},\\
&\qquad\quad \tilde G_n^{-1}(V_i^{(b)}-)+\frac{1}{n+1} \Big).
\end{align*}
where $\tilde F_n$ and $\tilde G_n$ are the empirical distribution functions
 of
$U_i^{(b)}$ and $V_i^{(b)}$ (and also of $U_i$ and $V_i$).
Note that
\begin{align*}
\overline{U}_{n:i}^{(b)}&=\overline{U}_{n:i}, \qquad
\underline{U}_{n:i}^{(b)}=\underline{U}_{n:i},\\
\overline{V}_{n:i}^{(b)}&=\overline{V}_{n:i}, \qquad
\underline{V}_{n:i}^{(b)}=\underline{V}_{n:i}.
\end{align*}
where the subscript of $A_{n:i}$ represents the $i$th order statistics
 (i.e., $i$th smallest number) of the sequence $\{A_i\}_{i=1}^n$.

We illustrate the tie-preserving procedure using the same toy example
with pseudo-observations~\eqref{eq:toy} in Section~\ref{sec:framework}.
The bootstrap pseudo-observations (without ties) after being sorted are always
\[
(U_{9:1}^{(b)},\ldots,U_{9:9}^{(b)}) = (1, 2, 3, 4, 5, 6, 7, 8, 9)/10.
\]
By applying~\eqref{eq:preserve_ties}, we obtain the tie-adjusted bootstrap
pseudo-observations
\[
(U_{9:1}^{(b)},\ldots,U_{9:9}^{(b)}) = \left(1,2,5,5,5,6,8,8,9 \right)/10,
\]
where we have changed $3/10$ and $4/10$ to $5/10$, and $7/10$ to $8/10$ to
match the ties in the observed pseudo-observations.
Consequently, the lower and upper boundaries of pseudo-observations of
$(U_{1}^{(b)}$, ... ,$U_{9}^{(b)})$ are
\begin{align*}
(\overline{U}_{9:1}^{(b)}, \ldots, \overline{U}_{9:9}^{(b)}) =
\left( 1,2,5,5,5,6,8,8,9 \right)/10,\\
(\underline{U}_{9:1}^{(b)}, \ldots, \underline{U}_{9:9}^{(b)}) =
\left( 1,2,3,3,3,6,7,7,9 \right)/10.
\end{align*}
The same procedure can be applied to the other margin $V_i$.

In summary, the tie-preserving parametric bootstrap procedure given the MPLE
$\hat\theta_n$ to construct a $1-\alpha$  confidence interval runs as
follows.
For some large integer $B$ , repeat the following steps (1) to (3) for
every $b \in  \{1,\ldots, B \}$:
\begin{enumerate}

\item
Generate bootstrap pseudo-observations with no ties
% $(U^{(b)}_1, V^{(b)}_1)$, ... ,$(U^{(b)}_n, V^{(b)}_n)$
from the fitted copula $C_{\hat{\theta}_n}$.

\item
Obtain tie-adjusted pseudo-observations
% $(U^{(b)}_1, V^{(b)}_1), \ldots, (U^{(b)}_n, V^{(b)}_n)$
via~\eqref{eq:preserve_ties}.

\item
Obtain the MPLE $\hat{\theta}^{(b)}_n$ using the tie-adjusted
pseudo-observations.
\end{enumerate}
A bootstrap sample
$(\hat{\theta}^{(1)}_n$, ... ,$\hat{\theta}^{(B)}_n)$
is formed to approximate the sampling distribution of $\hat\theta_n$.
The sample $\alpha/2$ and $1-\alpha/2$ quantiles can then be used to form
a confidence interval of level $1-\alpha$.

The computing cost of the tie-preserving parametric bootstrap procedure is
similar to that of the standard parametric bootstrap procedure.
The only extra part is the tie-preserving step, which is minimal compare to
the optimization in the fitting for each bootstrap sample.

\subsection{Goodness-of-Fit Test}
\label{sec:gof}

Goodness-of-fit tests with standard parametric bootstrap are known to be
vulnerable to ties in keeping their sizes \citep{Kojadinovic2010}.
This is because goodness-of-fit test statistics (usually distance-based) tend
to be bigger when ties are present.
However, when a standard parametric bootstrap generates tie-free samples,
it leads to under-estimation of the magnitude
of the null sampling distribution of the testing statistic.
Consequently, the tests would not hold their sizes with over rejection.
From our numerical studies, the empirical size of a 5\%-level test could be
100\% when even a moderate amount of ties are present.
Therefore, preserving ties in parametric bootstrap is crucial
\citep{Kojadinovic2016}.

We propose to adapt the standard bootstrap procedure for goodness-of-fit
\citep{Genest2008} with observed ties-preserved \citep{Kojadinovic2016}.
The null hypothesis is
\[
H_0: C \in \mathcal{C} = \{ C_\theta: \theta \in \Theta \} \quad
\textrm{versus} \quad
H_1: C \notin \mathcal{C}.
\]
Consider goodness-of-fit tests based on the goodness-of-fit empirical
process
\[
\begin{split}
\mathbb{C}_n ( u, v ) =&\sqrt{n} ( C_n ( u, v ) - C_{\hat{\theta}_n} ( u, v )),\\
&\quad ( u, v ) \in [ 0, 1 ]^2,
\end{split}
\]
where the $\emph{C}_n $ is the empirical copula defined as
\[
C_n ( u, v ) = \frac{1}{n} \sum_{i=1}^n { \mathbbm{1} ( U_i \le u, V_i \le v ) },
\]
and $\hat\theta_n$ is a parametric estimator of $\theta$
(which could be the MPLE from~\eqref{eq:mple} or other estimator)
under the null hypothesis $H_0$.
Statistics of goodness-of-fit tests can be formed as
$\mathcal{F}(\mathcal{C}_n)$, where
$\mathcal{F}$ is a functionals of $\mathbb{C}_n$.
Examples are Kolmogorov--Smirnov, Anderson--Darling, and
Cramer-von Mises (CvM) distance \citep{Genest2009, Kojadinovic2010}.
We use the CvM statistic, which has been known to have a good power
\citep{Kojadinovic2010}, to illustrate the procedure.

The CvM statistic is defined as
\begin{align}\label{eq:cvm}
D_n =& \int_{ [0, 1]^2 }
\mathbb{C}_n^2(u,v) \dif C_n(u,v)
\nonumber\\
=& \sum_{ i = 1 }^n \left( C_n ( \overline{U}_i, \overline{V}_i ) -
C_{\hat{\theta}_n} ( \overline{U}_i, \overline{V}_i ) \right).
\end{align}
After $D_n$ is obtained, we use the following bootstrap procedure to draw
samples from the distribution of $D_n$ under $H_0$.
For some large integer $B$, repeat the following steps for each
$b \in \{1,..., B \}$:
\begin{enumerate}
\item
Generate bootstrap pseudo-observations with no ties
% $(U^{(b)}_1, V^{(b)}_1), \ldots , (U^{(b)}_n, V^{(b)}_n)$
from the fitted copula $C_{\hat{\theta}_n}$.

\item
Obtain tie-adjusted pseudo-observations
% $(U^{(b)}_1, V^{(b)}_1), \ldots, (U^{(b)}_n, V^{(b)}_n)$
via~\eqref{eq:preserve_ties}.

\item
Obtain the MPLE $\hat{\theta}^{(b)}_n$ using the tie-adjusted
pseudo-observations.

\item
Obtain the empirical copula $C^{(b)}_n$
based on the tied-adjusted pseudo-observations.

\item
Obtain test statistic (CvM distance) $D^{(b)}_n$ using~\eqref{eq:cvm}.
\end{enumerate}
An approximated p-value of the observed test statistic is then
$\sum_{b=1}^{B}{\mathbbm{1}(D^{(b)}_n \ge D_n)}/B$.

Again, this tie-preserving bootstrap procedure has similar computing cost
compared to the standard parametric bootstrap procedure.
The difference from the procedure of \citet{Kojadinovic2016} is that, after
each tie-preserving bootstrap sample is obtained, we use the interval
censoring approach for estimation instead of average ranks.

\section{Numerical Studies}
\label{sec:simu}

A large-scale simulation study was carried out to assess the performance of
proposed methods in point estimation, interval estimation, and
goodness-of-fit.

\subsection{Point Estimation}
\label{subsec:poes}

We first study the accuracy of the point estimation of the proposed method
(denoted as ``censoring'') and compare it with two existing methods,
breaking ties at random (denoted as ``random'') and using the average of
ties (denoted as ``average'').
For the random method, we use the mean of 100 randomizations.
Data were generated from three one-parameter copulas
parameterized by Kendall's $\tau$, Clayton (C), Gumbel (G), and normal (N),
with $\tau \in \{0.1, \ldots, 0.9\}$ to control the dependence level.
Ties were introduced by rounding the first margin to the first decimal place.
Three sample sizes were considered $n \in \{100, 200, 400\}$.

\begin{figure*}[tbp] \centering
\includegraphics[width=\textwidth]{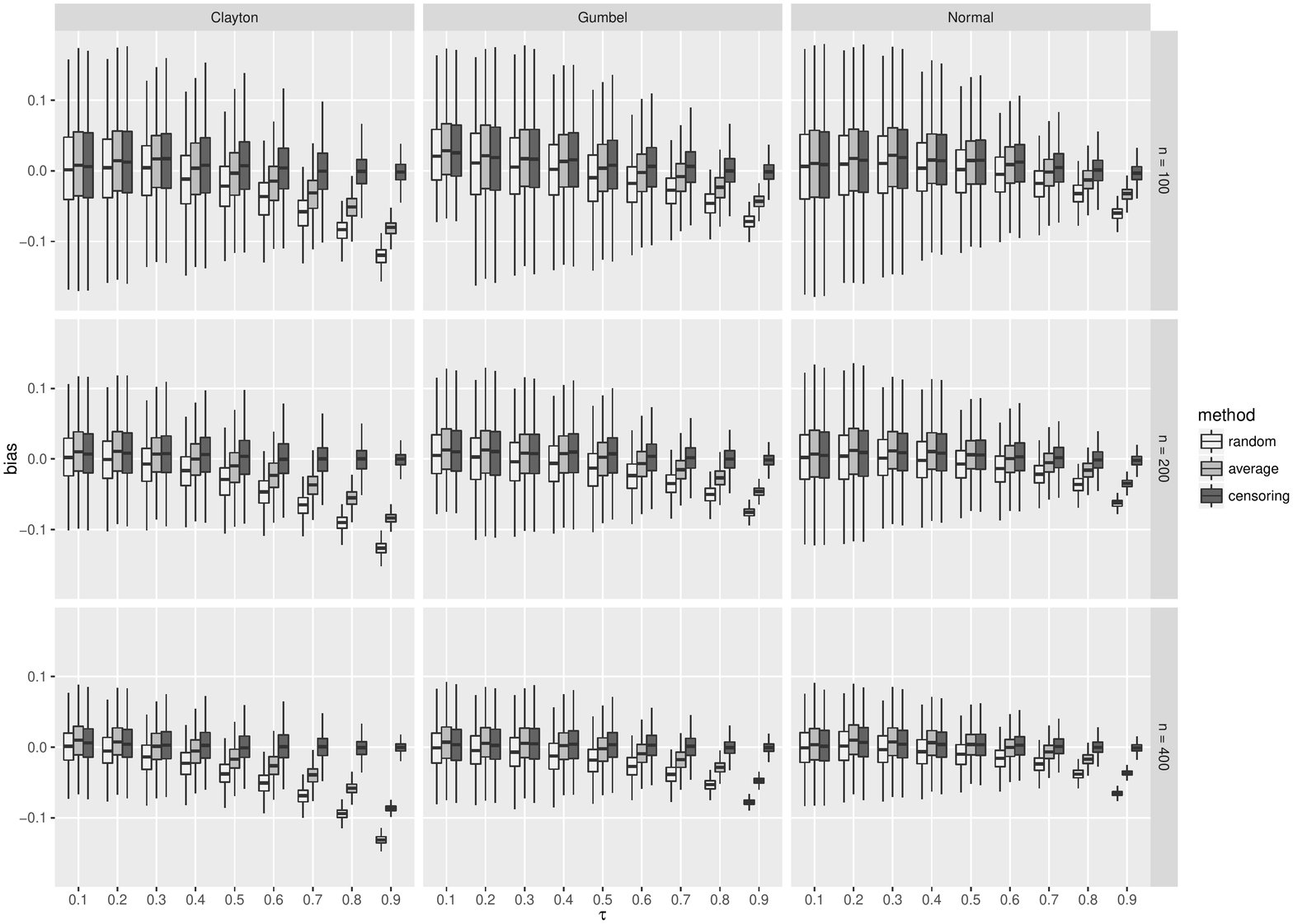}
\caption{
Boxplots of estimation error for Kendall's $\tau$ using three
methods (i.e., random, average, and censoring) for three types of copulas
(i.e., Clayton, Gumbel, and normal).
Sample size is $n \in \{100, 200, 400\}$.
Ties were introduced by rounding the first margin to the first decimal
place.}
\label{fig:points}
\end{figure*}

The estimation error of the MPLE estimator $\hat\tau_n$, $\hat{\tau}_n - \tau$,
from 1000 replicates are summarized in Figure~\ref{fig:points}.
It is clear that, as expected, the estimates from the average method and the
random method have little bias when the dependence is weak (lower $\tau$),
but as $\tau$ increases, they become more biased.
The estimate from the censoring method remains unbiased in all settings.
Variances of all three methods are comparable across all settings.
Therefore, the mean squared error (MSE) of the censoring method is smaller.
Furthermore, as the sample size increases, the variance of the estimate from the
censoring method reduces accordingly.

\begin{figure*}[tbp] \centering
\includegraphics[width=\textwidth]{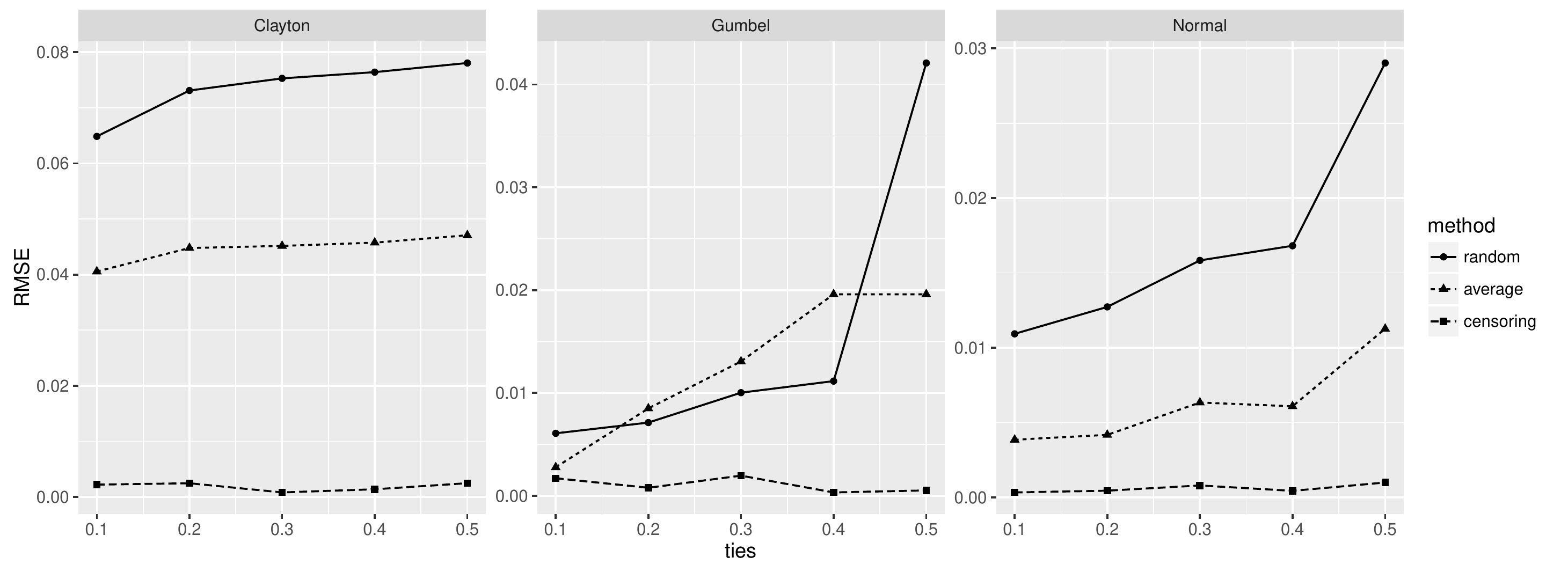}
\caption{Comparison of RMSEs of Kendall's $\tau$ for different methods
(i.e., random, average, and censoring) under three copulas
(i.e., Clayton, Gumbel and normal) with
different percentages of ties. Sample size is $n=200$.
Ties were introduced by rounding the first margin to the first decimal place.
}
\label{Fig:ties}
\end{figure*}

We then study the effect of the severity of ties on the estimation accuracy.
Data were generated from the three copulas with $\tau=0.75$ and $n=200$.
The first margin is rounded to the first decimal place if its value is
smaller than $\lambda$, which controls the percentage of ties.
We use the three methods to estimate $\tau$ and obtain their
corresponding root mean square errors (RMSEs) from 1000 replications.
These RMSEs are displayed in Figure \ref{Fig:ties}.
The censoring method has the smallest RMSE among the three methods, and
its RMSE remains stable regardless of the changes in the severity of ties.
The RMSEs of the average method and the random method increase as the
percentage of ties increases, with a faster rate for data generated from the
Gumbel copula.

\subsection{Interval Estimation}
\label{subsec:ines}

To assess the coverage properties of the bootstrap confidence intervals,
we generated data from the three copulas (C, G, and N) with Kendall's
$\tau \in \{0.25, 0.50, 0.75\}$ with sample size $n \in \{50, 100, 200\}$.
Ties were introduced by rounding the first margin to the first decimal place.
The 95\% confidence intervals of the censoring method were constructed with
the tie-preserving bootstrap procedure with bootstrap sample size $B = 1000$.

\begin{table*}[tbp]
 \caption {
Empirical coverage rate (in percentage) of the 95\% confidence interval of the
censoring method for different types of copulas (i.e., C=Clayton, G=Gumbel,
N=normal), different levels of Kendall's $\tau \in \{0.25, 0.5, 0.75\}$,
and different sample sizes $n \in \{50, 100, 200\}$.
Results are based on 500 replicates, each with bootstrap sample size $B = 1000$.
}
 \label{Tab:covRate}
 \centering
  \begin{tabular}{crrrrrrrrr}
    \toprule
    & \multicolumn{3}{c}{$\tau$ = 0.25}
    & \multicolumn{3}{c}{$\tau$ = 0.5} &
     \multicolumn{3}{c}{$\tau$ = 0.75} \\
     \cmidrule(lr){2-4} \cmidrule(lr){5-7} \cmidrule(lr){8-10}
    $n$ & C & G & N & C & G & N & C & G & N \\
    \midrule
    50  & 89.9 & 89.7 & 89.5 & 93.2 & 93.6 & 91.0 & 93.6 & 93.8 & 95.4 \\
    100 & 92.6& 92.4& 93.8& 95.4& 93.6& 92.6& 94.8& 96.2& 97.6 \\
    200 & 94.0 &93.4 &91.6 &96.0 &94.8 &93.0 &93.2 &	94.8 	&95.4 \\
    \bottomrule
  \end{tabular}
\end{table*}

The empirical coverage rates of the confidence intervals based on
500 replicates are summarized in Table~\ref{Tab:covRate}.
All the empirical coverage rates are close to the nominal level except that
in the setting with $n=50$ and $\tau=0.25$, the coverage rate is about 90\%.
The results suggest that the tie-preserving bootstrap procedure provides
confidence intervals that are valid for inferences for sample size over 100
or Kendall's $\tau$ over 0.50.

\subsection{Goodness-of-Fit Test}
\label{subsec:gof}

The finite-sample performance of goodness-of-fit tests
using the censoring method in estimation was assessed.
Data were generated from three copulas (C, G, and N) with Kendall's
$\tau \in \{0.25,0.5,0.75\}$ and sample size $n = 100$.
Three patterns of ties were considered: no ties or ties were introduced
by rounding one margin or both margins to the first decimal place.
For each configuration, we ran 500 replicates, for each replicate,
goodness-of-fit tests were performed with each of the three families of
copulas (C, G, and N) serving as the hypothesized copula.
The parametric bootstrap sample size was $B = 200$.
In the bootstrap procedure, two methods of preserving ties were considered:
matching the observed ranks as proposed in Section~\ref{sec:gof}, and the
rounding the margins with ties to the first decimal place.
Note that the rounding method is under that assumption of known
tie-introducing mechanism, which is unavailable in general.
We included this method as a benchmark only to investigate whether knowing
the tie-introducing mechanism helps to improve the performance of the tests.

\begin{table*}[tbp]
\caption{
Empirical rejection percentage of the goodness-of-fit tests with sample size
$n = 100$ for three types of copulas (C = Clayton, G = Gumbel, and N = Normal)
based on 500 replicates, each with bootstrap sample size $B = 200$.
Ties were introduced by rounding data from the first margin to first decimal
place.}
\label{Tab:gof100}
\centering
\begin{tabular}{lccc rrrrrr}
\toprule
  & & & & \multicolumn{6}{c}{Hypothesized copula}\\
  \cmidrule(lr){5-10}
  & Ties & Kendall's& True & \multicolumn{2}{c}{C} & \multicolumn{2}{c}{G}
  & \multicolumn{2}{c}{N} \\
  \cmidrule(lr){5-6} \cmidrule(lr){7-8} \cmidrule(lr){9-10}
  & pattern & $\tau$ & copula & Match & Round & Match & Round & Match & Round\\
     \midrule
     \\
     &No ties &0.25& C & \textbf{5.5} && 62.0 && 15.9 &\\
     &        &    & G & 79.1 && \textbf{5.3} && 16.8 &\\
     &        &    & N & 47.7 && 13.6  && \textbf{5.2} &
     \\[1ex]
     &        &0.5 & C & \textbf{7.0} && 96.8 && 60.0 &\\
     &        &    & G & 99.0 && \textbf{6.2} && 28.6 &\\
     &        &    & N & 88.2 && 24.0 && \textbf{5.0} &
     \\[1ex]
     &        &0.75& C & \textbf{4.0} && 100.0 && 85.8 &\\
     &        &    & G & 100.0 && \textbf{3.0} && 31.0 &\\
     &        &    & N & 98.2 && 21.6 && \textbf{3.0} &
     \\[3ex]
     &One side&0.25& C & \textbf{4.4} & \textbf{4.2}
     & 57.6 &  58.9 & 3.8  &18.2 \\
   	 & 		   & & G   & 76.5 & 75.9
   	 & \textbf{4.2} & \textbf{3.4} & 18.3& 18.6\\
   	 & 		  & & N   & 44.9 & 45.5 & 12.7 & 12.1
   	 & \textbf{3.5} & \textbf{6.4}
   	 \\[1ex]
     &		  &0.5 & C  & \textbf{4.6} & \textbf{5.4}
     & 95.4 & 95.6 & 52.2 & 55.2 \\
     &		  &	& G    & 99.8 & 99.6
     & \textbf{6.2} & \textbf{6.6} & 32.4 & 33.6 \\
     &		  &	 &  N   & 87.0 & 88.2 & 22.2 & 21.4
     & \textbf{3.6}& \textbf{4.2}
     \\[1ex]
     &		  &0.75& C & \textbf{1.6} & \textbf{3.8}
     & 99.6 & 99.6 &79.4 & 79.6\\
     &		  &	 & G   & 100  & 100
     & \textbf{4.0} & \textbf{4.2} & 25.6 & 26.6\\
     &		  &  & N   & 96.6 & 97.0 &14.4 & 15.4
     & \textbf{3.4}  & \textbf{3.8}
     \\[3ex]
     &Two sides &0.25&C &\textbf{6.4} & \textbf{4.4}
     & 51.4 & 55.6 & 13.8 & 15.4\\
     & 			&&G & 71.7 & 73.5
     & \textbf{4.7} & \textbf{3.7} & 19.4 &18.8\\
     & 			&&N & 40.2 & 38.6& 8.8 & 11.8
     & \textbf{5.0} & \textbf{4.2}
     \\[1ex]
     &			&0.5&C &\textbf{4.6} & \textbf{3.8}
     & 96.0 &  96.8 & 53.2 & 54.6 \\
     &			&&G & 98.6 & 99.0
     & \textbf{4.2} & \textbf{5.8}& 28.6 & 31.8\\
     &			&&N & 82.8 & 85.0 & 19.2 & 18.6
     & \textbf{5.6} & \textbf{4.8}
     \\[1ex]
     &			&0.75&C &\textbf{0.4} & \textbf{4.2}
     & 97.8 & 98.0 & 75.0 & 83.2 \\
     &			&&G & 99.6 &100.0 &
     \textbf{4.4} & \textbf{5.4}& 26.6 &30.6 \\
     &			&&N & 93.6 & 94.8 & 10.6 & 15.0
     & \textbf{4.6} & \textbf{4.4}
     \\
	 \bottomrule
\end{tabular}
\end{table*}

The empirical rejection percentages of the goodness-of-fit tests
with level 5\% are summarized in Table~\ref{Tab:gof100}.
When the hypothesized copula is the same as the data generating copula, the
reported percentages are put in bold, representing the empirical sizes.
The empirical sizes are close to the nominal size of 5\% in most cases.
The two methods of preserving ties showed little difference, except that the
test is conservative for the Clayton copula with $\tau = 0.75$, with
empirical rejection percentage 1.6 and 0.4, respectively, for one and two
side ties. When the hypothesized copula is not the data generating copula,
the empirical powers of the tests are lower than those obtained when no
ties are present. This is expected due to the information loss in ties.
Between the two tie-perserving methods, the rounding approach seems
to have slightly higher power, but the advantage seems quite limited.
Note that, however, the rounding approach may not be applicable in practice
because we may not know the true tie-introducing mechanism.

Now that the difference between the two tie-preserving methods is little, we
focus on the matching ties method and investigate sample sizes 50 and 200.
The results are summarized in Table~\ref{Tab:gofMore}.
As expected, the test holds its size better at sample size 200, and the
power increases as the sample size increases in all settings.
% The results in Kojadinovic2016 cannot be directly compared!

\begin{table*}[tbp]
\caption{
Empirical rejection percentages of the goodness-of-fit tests
with sample size $n \in \{50, 200\}$ for three types of copulas
(C = Clayton, G = Gumbel, and N = Normal)
based on 500 replicates, each with bootstrap sample size $B = 200$.
Ties were introduced by rounding data from the first margin to first decimal
place. Matching rank was used to preserve ties in bootstrap sample.}
\label{Tab:gofMore}
\centering
\begin{tabular}{lccc rrrrrr}

\toprule
  & & & & \multicolumn{6}{c}{Hypothesized copula}\\
  \cmidrule(lr){5-10}
  & Ties & Kendall's& True & \multicolumn{2}{c}{C} & \multicolumn{2}{c}{G}
  & \multicolumn{2}{c}{N} \\
  \cmidrule(lr){5-6} \cmidrule(lr){7-8} \cmidrule(lr){9-10}
  & pattern & $\tau$ & copula &
  $n=50$ & $n=200$ & $n=50$ & $n=200$ & $n=50$ & $n=200$\\
     \midrule
     \\
     \\
     &One side&0.25& C & \textbf{4.3} & \textbf{6.4}
     & 32.5 & 88.8 & 6.0 &41.8\\
     &        &    & G & 50.7 & 94.2 & \textbf{6.9}
     & \textbf{5.4} & 9.4 &30.4\\
     &        &    & N & 24.1 & 69.4 & 10.7 & 25.6
     & \textbf{2.6} & \textbf{6.8}
     \\[1ex]
     &        &0.5 & C & \textbf{3.2} & \textbf{4.0}
     & 72.9 & 100.0 & 21.9 &93.8\\
     &        &    & G & 89.2 & 100.0 & \textbf{4.7}
     &\textbf{5.2} & 21.8 &42.2\\
     &        &    & N & 59.2 & 99.8 & 11.6 & 38.6 &
     \textbf{5.0} & \textbf{4.6}
     \\[1ex]
     &        &0.75& C & \textbf{1.1} & \textbf{5.2}
     & 83.7 &100.0 & 35.1 & 99.0 \\
     &        &    & G & 94.6 & 100.0 & \textbf{2.9}
     &\textbf{5.8}& 17.1 &47.6\\
     &        &    & N & 73.3 & 100.0 & 7.4 & 32.2
     & \textbf{3.9} & \textbf{4.0}
     \\[1ex]
     &Two sides&0.25& C & \textbf{4.6} & \textbf{4.4}
     & 11.9 &55.6 & 1.5 &18.0\\
     &         &    & G & 46.5 & 73.5 & \textbf{3.3}
     & \textbf{3.7} & 10.8 & 19.2\\
     &         &    & N & 24.2 & 38.6 & 6.8 & 11.8
     & \textbf{3.9} & \textbf{4.8}
     \\[1ex]
     &         &0.5 & C & \textbf{2.9} & \textbf{3.8}
     & 53.2 & 96.8& 8.4 & 55.0\\
     &         &    & G & 86.0 & 99.0 & \textbf{2.5}
     & \textbf{5.8} & 16.1 & 26.4\\
     &         &    & N & 57.2 & 85.0 & 10.3 & 18.6
     & \textbf{3.4} & \textbf{4.2}
     \\[1ex]
     &         &0.75& C & \textbf{0.7} & \textbf{4.2}
     & 75.9 & 98.0 & 34.6 &78.4 \\
     &         &    & G & 89.8 & 100.0 & \textbf{1.0}
     & \textbf{5.4} & 13.9 & 30.6\\
     &         &    & N & 60.2 & 94.8 & 2.8 & 15.0
     & \textbf{1.8} & \textbf{5.0}
     \\
   \bottomrule
\end{tabular}
\end{table*}

\section{Real Data Example}
\label{sec:reals}

The bivariate insurance data considered in \citet{FreesValdez1998} has
often been used as illustration in copula modeling \citep{Kojadinovic2010}.
The two variables are indemnity payment and allocated loss adjustment
expense, observed from 1466 uncensored claims of an insurance company.
A lot of ties are present in indemnity payment, with only 541 unique values.
Ties are much less in allocated loss adjustment expense (1401 unique values).
Existing works have demonstrated that it is necessary to account for ties
to analyze this data set \citep{Kojadinovic2016}.

We performed goodness-of-fit tests for four copulas, Clayton, survival
Clayton, Gumbel, and normal, using the censoring method with the
tie-preserving bootstrap procedure with bootstrap sample size $B = 1000$.
The p-values for Clayton, survial Clayton, Gumbel, and normal copulas are
$0.000$, $0.000$, $0.168$, and $0.000$, respectively.
Only the Gumbel copula is not rejected at the 5\% level, which is consistent
with results in existing studies \citep{Kojadinovic2010, Kojadinovic2016}.
In particular, \citet{Kojadinovic2016} tested goodness-of-fit for the same four
copula families with ties taken into account, and reported that the Gumbel
copula was the only one not rejected at the 1\% level, with p-value $0.0225$.
The difference between our p-values and theirs may be due to the estimation
methods: \citet{Kojadinovic2016} used the average rank method while we
used the interval censoring method.

Assuming that the true copula is a Gumbel Copula, the parameter
estimate from the interval censoring method is $1.425$, with 95\%
tie-preserving bootstrap confidence interval $(1.366, 1.505)$.
The point and interval estimate from the averaging the tied ranks are
$1.424$ and $(1.362, 1.498)$.
The difference between the two methods is small in this example.
This may be explained by the moderate dependence in the data.
The MPLE of Kendall's $\tau$ is $0.298$, which is close to the empirical
Kendall's $\tau=0.309$ calculated in presence of ties.

\section{Discussion}
\label{sec:disc}

Unlike the average rank approach, independence randomization
\citep{Kojadinovic2010}, or co-monotone/mixed randomization
\citep{Pappada2016}, the interval censoring approach does not
distort the features of the observed data.
Consequently, it does not have the bias that other approaches may
have introduced, especially when the dependence is strong.
When the dependence is weak, although the point estimate may not be very
different from the point estimate with the average rank method, the small
difference might still propogate to become important when estimation is
repeatitively needed as in the case of parametric bootstrap procedures.
The interval censoring method can be applied to model discrete data,
in which case it has the same spirit as \citet{Aristidis2009}.
The limiting distribution of the MPLE using the interval censored
pseudo-observations is a challenging problem for two reasons.
First, likelihood estimator from interval censored data do not achieve the
standard $n^{1/2}$-rate \citep{wellner1995interval, van2000preservation}.
Second, the interval censored data used in the estimation are
pseudo-observations resulting from the probability integral transform with
marginal empirical distribution functions, instead of the observations.
Establishing the asymptotic properties of the MPLE from interval-censored
pseudo-observations would be a contribution of strong interest.

The tie-preserving parametric bootstrap procedure provides valid finite
sample inferences for the estimator from the interval censoring method.
The procedure can be applied to many inference problems
for copula modeling with tied data \citep{Kojadinovic2016}.
The parameter estimation step in the procedure for bootstrap sample with
ties could use the average rank method as in \citet{Kojadinovic2016},
which would, however, leads to biased estimation with strong dependence.
A combination of the interval censoring method for estimation and the
tie-preserving bootstrap procedure for inference appears to be a practical
approach to rank-based copula modeling for data with ties.
Applications to inferences such as tests for exchangeability, extreme-value
dependence, and radial symmetry merits further research.

\begin{acknowledgements}
J. Yan's research was partially supported by an NSF grant (DMS 1521730).
Yang Li's research was partially supported by the Fundamental Research Funds
for the Central Universities, and the Research Funds (15XNI011) of
Renmin University of China.
\end{acknowledgements}

\bibliographystyle{Chicago}
\bibliography{copties}

\end{document}